\input miguel.sty

\vglue2.5truecm

\titulo{On the Topology of the Symmetry Group of the Standard Model}

\bigskip

\hangindent=1.5cm
{\bf M.$\!$ A. Aguilar\footnote{$^{1}$}{Instituto de
Matem\'aticas, Universidad Nacional Aut\'onoma de M\'exico, Circuito
Exterior, Cd.  Universitaria, 04510 M\'exico, D.$\!$~F.; e-mail:
marcelo@math.unam.mx} and M. Socolovsky\footnote{$^{2}$}{Instituto de
Ciencias Nucleares, Universidad Nacional Aut\'onoma de M\'exico,
Circuito Exterior, Cd. Universitaria, 04510 M\'exico,
D.$\!$~F., e-mail: socolovs@nuclecu.unam.mx}}

\bigskip

\bigskip

\hangindent=1.5cm {\tenpoint
\hskip.8cm
We study the topological structure of the symmetry group of the
standard model, $G_{SM}=U(1)\times SU(2)\times SU(3)$. Locally,
$G_{SM}\cong S^1\times (S^3)^2\times S^5$. For $SU(3)$, which is an
$S^3$-bundle over $S^5$ (and therefore a local product of these
spheres) we give a canonical gauge {\it i.e.\/} a canonical set of
local trivializations. These formulae give the matrices of
$SU(3)$ in terms of points of spheres. Globally, we prove that the
characteristic function of $SU(3)$ is the suspension of the Hopf map
$S^3 \aderecha{h} S^2$. We also study the case of $SU(n)$ for 
arbitrary $n$, in particular the cases of $SU(4)$, a flavour 
group, and of $SU(5)$, a candidate group 
for grand unification. We show that the $2$-sphere is also related to
the fundamental symmetries of nature due to its relation to
$SO^0(3,1)$, the identity component of the Lorentz group, a 
subgroup of the symmetry group of several gauge theories of 
gravity.}

\bigskip

\section{1. Introduction} 

\

As is well known, the symmetry group of the electroweak and strong forces 
(standard model) before spontaneous symmetry breaking is given by
(Taylor, 1976)
$$G_{SM}=U(1)\times SU(2)\times SU(3)$$
After symmetry breaking, however, $G_{SM}$ breaks down to
$G^\prime_{SM}=U(1)\times SU(3)$, the remaining exact symmetry of the
electromagnetic and color forces. In any case, $U(1)$ is the circle
or {\it $1$-sphere\/} $S^1$, the unit complex numbers; while
$SU(2)$, given by all complex matrices $A=\pmatrix{z & w\cr -\bar
w&\bar z\cr}$ with $\det A=1$ is the {\it $3$-sphere\/} $S^3$ or unit 
quaternions, since if $z=\alpha +i\beta$ and $w=\gamma
+i\delta$ then the condition of unit determinant is
$\alpha^2+\beta^2+\gamma^2+\delta^2=1$. 

\

In the mathematical literature it is well known that for all
$n=2,3,\ldots$ the groups $SU(n)$ are principal $SU(n-1)$-bundles over the
$(2n-1)$-spheres {\it i.e.\/} that one has pairs of maps
(Steenrod, 1951)
$$SU(n-1)\ \aderecha{\iota}\ SU(n)\ \aderecha{\pi_n}\ S^{2n-1}$$
where $\iota$ is the canonical inclusion and $\pi_n$ maps a matrix $A$ to 
$Ae_0$, where $e_0$ is the vector whose entries are all 0 except the last 
one which is 1. In particular for $n=3$ one has the $SU(2)$-bundle
$$SU(2)\to SU(3)\aderecha{\pi_3} S^5$$
which in particular means that locally $SU(3)\cong S^5\times S^3$
since $SU(2)\cong S^3$; moreover, according to the theory of bundles, 
the isomorphism classes of 
$SU(2)$-bundles over $S^5$ are in one-to-one correspondence with the
$4$-th homotopy group of $SU(2)$ {\it i.e.\/} $
k_{SU(2)}(S^5)\leftrightarrow \Pi_4(SU(2))\cong \Pi_4(S^3)\cong
\Z_2=\{0,1\}$: $0$ corresponds to the trivial bundle, $S^5\times S^3$ 
while $1$ corresponds to $SU(3)$ (see Section 3.2). 
In other words, $SU(3)$, the symmetry group
of the strong interactions, {\it is the unique\/} (up to isomorphism)
{\it non-trivial\/ $SU(2)$-bundle over the $5$-sphere}, and as this
result shows, it is also constructed from spheres, though not
globally. This means that
$$G_{SM}\mathop{=}\limits_{\rm loc.}S^1\times (S^3)^2\times S^5$$ 
and, after symmetry breaking, $G^\prime_{SM}\mathop{=}\limits_{\rm
loc.}S^1\times S^3\times S^5$.

\

For higher $n$, however, uniqueness is lost since, for example,
for $n=4$ and $n=5$ one has the bundles
$$SU(3)\to SU(4)\aderecha{\pi_4} S^7$$
and
$$SU(4)\to SU(5)\aderecha{\pi_5} S^9$$
respectively, and $k_{SU(3)}(S^7)\leftrightarrow  
\Pi_6(SU(3))\cong \Z_6$ and $k_{SU(4)}(S^9)\leftrightarrow \Pi_8(SU(4))\cong
\Z_{24}$ (EDM, 1993). Notice however that locally any $SU(n)$ is a
topological product of odd-dimensional spheres:
$SU(4)\mathop{=}\limits_{\rm loc.}S^7\times
SU(3)\mathop{=}\limits_{\rm loc.}S^7\times S^5\times S^3$,
$SU(5)\mathop{=}\limits_{\rm loc.}
S^9\times SU(4)\mathop{=}\limits_{\rm loc.}S^9\times S^7\times
S^5\times S^3,\ldots,SU(n)\mathop{=}\limits_{\rm loc.}S^{2n-1}\times
S^{2n-3}\times \cdots \times S^5\times S^3$. This expression allows
us to define a formula which gives, in a canonical way, any element of
$SU(n)$ in terms of points of spheres. 

\

It is interesting to remark here that a typical gauge theory, say on
Minkowski space-time $M^4$, is a theory of connections on the
trivial bundle $M^4\times G$, where $G$ is the symmetry group,
coupled to sections of associated bundles; in the case of the strong and
weak interactions the groups themselves are principal bundles: for
the weak case, $SU(2)$ is the total space of the Hopf bundle $S^1\to
S^3\aderecha{\kappa_2} S^2$.

\

In Section~2 we briefly review the global construction of the bundle
$SU(2)\to SU(3)\aderecha{\pi_3} S^5$ and construct a canonical set of local
trivializations of $SU(3)$, starting from the (canonical) homogeneous
coordinates on $\C P^2$, the complex projective plane. These formulae
exhibit $SU(3)$ as a local product of spheres and moreover, give
explicit expressions for all the matrices of $SU(3)$ in terms of the
spheres $S^5$ and $S^3$.

\

The above choice of coordinates is natural since for all $n\geq 2$,  
$S^{2n-1}$ is a principal bundle over $\C P^{n-1}$ with fiber $S^1$ (complex 
Hopf bundles):
$$\matrix{ & & & & S^1\cr
 & & & & \downarrow \cr
 SU(n-1) & \to & SU(n) & \aderecha{\pi_n} & S^{2n-1} \cr
  & & & & \downarrow^{\kappa_n} \cr
  & & & & \C P^{n-1} \cr}$$
and $\C P^{n-1}$ has $n$ canonical charts defining its homogeneous 
coordinates. Then the bundle $\pi_n$, for all $n$, can be locally trivialized in a 
canonical way, $n$ being the number of local trivializations. In the 
following the Hopf map $\kappa_2$ will be denoted by $h$. If 
$\pmatrix{z\cr w\cr} \in S^3 \subset \C^2$ $(\vert z \vert^2+\vert w 
\vert^2=1)$ and $\C P^1$ is identified with the Riemann sphere $\hat{\C}
= \C \cup \{\infty \}$ then $h$ is given by 
$$
h \pmatrix{z\cr w\cr}=
\bigg\{ \matrix{z/w, w \neq 0 \cr \infty, w=0. \cr}
$$
It can be proved that $h$ is {\it essential i.e.} it is not homotopic 
to a constant map (Spanier, 1966). 

\

In Section~3 we prove that the characteristic (or clutching) map of 
$SU(3)$ is the suspension of the Hopf map $S^3 \aderecha{h} S^2$. Since 
the clutching map allows to construct the bundle, then $SU(3)$ is built 
from information contained in the Hopf map. This map, besides having 
great importance in homotopy theory, plays a relevant r\^ole in physics 
e.g. in the geometrical description of the spin $1 \over 2$ system 
(Ashtekar and Schilling, 1994; Corichi and Ryan, 1997), and the Dirac 
monopole of unit magnetic charge (Wu and Yang, 1975). 

\

In Section 4 we investigate the general case of 
$SU(n)$ and, using a result of Steenrod for $U(n)$, we 
prove that for even $n$, $n \geq 2$, the characteristic map $g_{n+1}:
S^{2n}\to SU(n)$ 
of $SU(n+1)$ is a homotopy lifting of the $(2n-3)-th$ suspension of $h$. 
(If $Z\aderecha{p} Y$ is a projection and $X\aderecha{f} Y$ is a 
continuous function, then $p$ lifts $f$ if there is a continuous function 
$X\aderecha{g} Z$ such that $p\circ g=f$. The lifting is up to homotopy 
if $p\circ g \sim f$.) 
The case of $SU(5)$ is interesting since it is a candidate group for
grand unification (Mohapatra, 1986). On the other hand, for odd $n$, 
$n \geq 3$, $\pi _n \circ g_{n+1}$ is inessential. A particular case is 
$SU(4)$, which is a flavour group.

\

In Section 5 we briefly discuss how the {\it $2$-sphere\/} $S^2$
appears in the context of the symmetry group of the fundamental
interactions, due to the canonical isomorphism between its conformal
group and $SO^0(3,1)$, 
the proper orthochronous Lorentz group.

\section{2. The bundle $S^3\to SU(3)\aderecha{\pi_3} S^5$}

\subsection{2.1. The groups $U(3)$ and $SU(3)$} 

\

The $n$-dimensional complex vector space $\C^n$ equipped with the
Hermitian scalar product $\langle \vec{z},\vec{w}\rangle
=\sum\limits^n_{i=1}\bar z_iw_i$ is a Hilbert space. The
$n\times n$ complex matrices which leave 
$\langle ,\rangle$ invariant form the group $U(n)$ {\it i.e.\/}
$U(n)=\Aut (\C^n,\langle ,\rangle )$: {\it the group of automorphisms 
of\/ $\C^n$ as a Hilbert space}. If $A\in U(n)$ and $A^*$ is the
transpose conjugate matrix, then $A^*A=I$ {\it i.e.\/} $A^*=A^{-1}$,
so $\vert \det A\vert =1$ and $\dim_{\R}U(n)=n^2$. The topology of
$U(n)$ is inherited from the vector space of $n\times n$ complex 
matrices, which is isomorphic to Euclidean space $E^{2n^2}$. $U(n)$ 
is a Lie group and $SU(n)$ is the closed Lie subgroup consisting of 
matrices whose determinant is 1. Since $U(n)$ is compact, 
$SU(n)$ is also compact. 

\

For $n=3$, $SU(3)$ is 2-connected {\it i.e.} 
$\Pi_k(SU(3))=0$ for $k=1,2$, and $\Pi_3(SU(3))\cong \Z)$.
Topologically, $U(3)\cong SU(3)\times U(1)$, so $U(3)$ is 
connected but not $1$-connected since $\Pi_1(U(3))\cong \Z$. 

\subsection{2.2. The inclusion and action $SU(2)\to SU(3)$} 

Let $\iota \colon SU(2)\to SU(3)$, $\pmatrix{z&w\cr -\bar w&\bar
z\cr}\mapsto \pmatrix{z&w&0\cr -\bar w&\bar z&0\cr 0&0&1\cr}$ be the
inclusion of $SU(2)$ into $SU(3)$, call $SU(2)^\prime =\iota (SU(2))$.
Clearly $SU(2)^\prime \cong SU(2)$, both topologically and as a
group. The right action $SU(3)\times SU(2)^\prime \to SU(3)$ is given by
matrix multiplication $(B,A)\mapsto BA$ {\it i.e.}
$$\pmatrix{\alpha &\beta &\gamma\cr \sigma &\varepsilon &\varphi \cr
\kappa &\lambda &\mu\cr}\pmatrix{z&w&0\cr -\bar w&\bar z&0\cr 0&0&1\cr}=
\pmatrix{
\alpha z-\beta \bar w       &\alpha w+\beta \bar z       &\gamma\cr 
\delta z-\varepsilon \bar w &\delta w+\varepsilon \bar z &\varphi\cr 
\kappa z-\lambda \bar w           &\kappa w+\lambda \bar z    &\mu\cr}$$

Let $q\colon SU(3)\to SU(3)/SU(2)^\prime$ be the quotient map, {\it i.e.} 
$q(B)=[B]$, 
where $SU(3)/SU(2)^\prime$ is the orbit space $\{[B]\}_{B\in
SU(3)^\prime}$ with the quotient topology, and 
$[B]=BSU(2)^\prime$, in particular 
$[I]=SU(2)^\prime$. Notice that $\vert \gamma \vert^2+
\vert \varphi \vert^2+\vert \mu \vert^2=1$ {\it i.e.\/} $\pmatrix{\gamma\cr
\varphi\cr \mu\cr}\in S^5\subset \C^3$. It is easy to verify that the
following diagram commutes:
$$\matrix{  & SU(3)  & \cr  {}^q\swarrow  &  & \searrow^{\pi_3}  \cr
SU(3)/SU(2)^\prime & \aderecha{\kappa}\ & \ S^5 \cr}$$
where $\pi_3(B)=\pmatrix{\gamma \cr \varphi \cr \mu\cr}$ and
$\kappa ([B])=B\pmatrix{0\cr 0\cr 1\cr}$, in particular
$\kappa ([I])=\pmatrix{0\cr 0\cr 1\cr}$; $\kappa $ turns out to be a {\it 
homeomorphism\/} with inverse $\kappa^{-1}\pmatrix{\gamma\cr \varphi\cr
\mu\cr}=B^\prime SU(2)^\prime$ for any $B^\prime =\pmatrix{\cdots
&\gamma\cr \cdots &\varphi\cr \cdots &\mu\cr}\in SU(3)$ (an explicit
formula for $\kappa^{-1}$ will be given in Section~2.3). Clearly
$\pi_3^{-1}\left(\!\left\{\!\pmatrix{\gamma\cr \varphi\cr
\mu\cr}\!\right\}\!\right)\!=B^\prime SU(2)^\prime \cong SU(2)^\prime
\cong SU(2)\cong S^3$, so $S^3$ is the fiber of $\pi_3$.

\subsection{2.3. Local trivializations} 

\

Consider the $S^1$-bundle $S^5\ \aderecha{\kappa_3}\ \C P^2$, where the
complex projective plane is the space of complex lines through the
origin in $\C^3$; $\C P^2$ has three canonical charts given by the
open sets $V_k=\{\vec{z}(\C \setminus \{0\})\mid \
\vec{z}=\pmatrix{z_1\cr z_2\cr z_3\cr}$, and 
$\ z_k\ne 0\} \subset \C P^2$, for $k=1, 2, 3$, and
the homeomorphisms $V_k\to \C^2$ map $\vec{z}(\C \setminus \{0\})$ to 
$(\xi_i,\xi_j)=(z_i/z_k,z_j/z_k)$ with $i,j,k$ in cyclic order, the 
$\xi_i$ are called homogeneous coordinates. Then the pre-images of $V_k$
by the projection $\kappa_3$ define three open sets in $S^5$ given by
$U_k=\kappa_3^{-1}(V_k)=\{\pmatrix{z_1\cr z_2\cr z_3\cr}
\mid z_k\ne 0\}\equiv
S^5_k \subset S^5$; $\bigcup\limits^3_{i=1}U_i=S^5$
with $U_i\cap U_j\ne \phi$ for all $i,j$ and $(0,0,1)\in U_3 \subset S^5$
but $(0,0,1)\notin U_1,U_2$. Notice that the complements of $S^5_k$
with respect to $S^5$ are homeomorphic to $S^3\colon
(S^5_3)^c=S^5\setminus S^5_3=\{\pmatrix{z_1\cr z_2\cr 0\cr}
\mid \vert z_1\vert^2
+\vert z_2\vert^2=1\}\cong S^3$, and analogous formulae for $S^5_1$
and $S^5_2$; $(S^5_k)^c$ are closed sets in $S^5$. We shall
trivialize the bundle $SU(3)\aderecha{\pi_3} S^5$ over the $U_k$'s.

\

In order to construct local sections of the bundle $\pi _3$, consider the 
following complex matrices:
$$\tilde{C}=\pmatrix{1&0&a\cr 0&1&b\cr 0&0&c\cr},\qquad
\tilde{B}=\pmatrix{1&0&a\cr 0&0&b\cr 0&1&c\cr},\qquad \hbox{ and }
\tilde{A}=\pmatrix{0&0&a\cr 1&0&b\cr 0&1&c\cr}$$
It is easy to verify that the three column vectors in each of them 
are linearly independent if, respectively, $c\ne 0$, $b\ne 0$ and
$a\ne 0$. In the three cases we take $\vert a\vert^2+\vert
b\vert^2+\vert c\vert^2=1$. By the Gram-Schmidt procedure we can
construct unitary matrices $\hat{C}$, $\hat{B}$, and $\hat{A}$ and
then, multiplying each of them from the right by the matrix  
$\pmatrix{z & 0 \cr 0 & I \cr}$, where $z^{-1}=det\hat{C}$, $det\hat{B}$, 
or $det\hat{A}$, we obtain the following elements of $SU(3)$: 
$$C=\pmatrix{
{\vert c\vert^2 \over c \sqrt{1-\vert b\vert^2}} &{-a\bar b\over
\sqrt{1-\vert b\vert^2}} &a\cr
\noalign{\vskip3pt}
0 &\sqrt{1-\vert b\vert^2} &b\cr
\noalign{\vskip3pt}
{-\bar a\over \sqrt{1-\vert b\vert^2}} &{-\bar bc\over 
\sqrt{1-\vert b\vert^2}} &c\cr}$$
$$B=\pmatrix{
{-\vert b\vert^2 \over b \sqrt{\vert a\vert^2+\vert b\vert^2}} &{-a\bar
c\over \sqrt{1-\vert c\vert^2}} &a\cr
\noalign{\vskip3pt}
{-\bar a\over \sqrt{\vert a\vert^2+\vert b\vert^2}}
&{-b\bar c\over \sqrt{1-\vert c\vert^2}} &b\cr 
\noalign{\vskip3pt}
0 &\sqrt{1-\vert c\vert^2} &c\cr},\qquad
A=\pmatrix{
{-\bar b\over \sqrt{1-\vert c\vert^2}} &{-a\bar c\over 
\sqrt{1-\vert c\vert^2}} &a\cr
\noalign{\vskip3pt}
{\vert a\vert^2 \over a \sqrt{1-\vert c\vert^2}} &{-b\bar c\over
\sqrt{1-\vert c\vert^2}} &b\cr 
\noalign{\vskip3pt}
0 &\sqrt{1-\vert c\vert^2} &c\cr}$$
with $det\hat{C}=c/\vert c\vert$, $det\hat{B}=-b/\vert b\vert$ and $det\hat{A}=a/\vert a\vert$.

\

(These formulae give an explicit 
expression for the map $\kappa^{-1}$ of Section~2.2: given
$\pmatrix{\gamma\cr \varphi\cr \mu\cr}\in S^5$, we choose $B^\prime$
equal to $A$, $B$ or $C$ if, respectively, $\gamma$, $\varphi$ or
$\mu$ is $\ne 0$.) We define local sections $\sigma_k\colon S^5_k\to
SU(3)$ as follows:

$\sigma_1\colon S^5_1\to SU(3)$, $\pmatrix{a\cr b\cr c\cr}\mapsto
\sigma_1\pmatrix{a\cr b\cr c\cr}=A$,

$\sigma_2\colon S^5_2\to SU(3)$, $\pmatrix{a\cr b\cr c\cr}\mapsto
\sigma_2\pmatrix{a\cr b\cr c\cr}=B$,

and

$\sigma_3\colon S^5_3\to SU(3)$, $\pmatrix{a\cr b\cr c\cr}\mapsto
\sigma_3\pmatrix{a\cr b\cr c\cr}=C$.

If $\pi \colon P\to X$ is a principal $G$-bundle over $X$, 
and $\sigma_\beta \colon U_\beta \to P$ are local sections,
then $\varphi_\beta \colon \pi^{-1}(U_\beta)\equiv P_\beta \to U_\beta
\times G$, where $\varphi_\beta (p)=(x,\gamma_\beta (p))$ with $x=\pi (p)$
and $p=\sigma_\beta (\pi (p))\cdot \gamma_\beta (p)$, are local
trivializations. If $\varphi_\alpha$ and $\varphi_\beta$ are local
trivializations and $U_\alpha \cap U_\beta \ne \phi$, then
$\varphi_\beta \circ \varphi^{-1}_\alpha \colon (U_\alpha \cap
U_\beta )\times G\to (U_\alpha \cap U_\beta) \times G$ satisfies 
$\varphi_\beta \circ \varphi^{-1}_\alpha (x,g)
=(x,g_{\beta \alpha}(x))$, 
where $g_{\beta \alpha}:U_\alpha \cap U_\beta \to G$ are the {\it 
transition functions} and $\sigma _\beta (x) \cdot g_{\beta \alpha}(x)
=\sigma_\alpha (x)$. In our case, with 
$G=SU(2)$, $P=SU(3)$, $X=S^5$, $\beta =k=1,2,3$, $U_k=S^5_k$, and
$P_k=SU(3)_k=\left\{\pmatrix{\cdots &z_1\cr \cdots &z_2\cr \cdots
&z_3\cr}\in SU(3)\mid z_k\ne 0\right\}$, the local trivilizations of
the $SU(2)$-bundle $p\equiv \pi_3 \colon SU(3)\to S^5$ are:

$\varphi_1\colon SU(3)_1\to S^5_1\times SU(2)^\prime$,
$\varphi_1(R)=(p(R),(\sigma_1(p(R)))^{-1}R)=\left(\pmatrix{a\cr b\cr
c\cr},A^*R\right)$, 

$\varphi_2\colon SU(3)_2\to S^5_2\times SU(2)^\prime$,
$\varphi_2(S)=(p(S),(\sigma_2(p(S)))^{-1}S)=\left(\pmatrix{a\cr b\cr
c\cr},B^*S\right)$, 

and

$\varphi_3\colon SU(3)_3\to S^5_3\times SU(2)^\prime$,
$\varphi_3(T)=(p(T),(\sigma_3(p(T)))^{-1}T)=\left(\pmatrix{a\cr b\cr
c\cr},C^*T\right)$.

The matrices $A^*R$, $B^*S$ and $C^*T$ are of the form $\pmatrix{
  &  &0 \cr
  &  &0 \cr
0 &0 &1 \cr}$\hglue-47.5truept \raise6pt\hbox{\bigggfntit D}\hglue47.5truept 
with $D\in SU(2)$. $\varphi_1$, $\varphi_2$ and $\varphi_3$ exhibit
the local structure of $SU(3)$. The transition functions are:

$g_{12}\colon S^5_1\cap S^5_2\to SU(2)^\prime$, $g_{12}\pmatrix{a\cr
b\cr c\cr}=A^*B$,

$g_{23}\colon S^5_2\cap S^5_3\to SU(2)^\prime$, $g_{23}\pmatrix{a\cr
b\cr c\cr}=B^*C$,

and

$g_{31}\colon S^5_3\cap S^5_1\to SU(2)^\prime$, $g_{31}\pmatrix{a\cr
b\cr c\cr}=C^*A$.

Notice that $I\in SU(3)_3$ but $I\notin SU(3)_j$ for $j=1,2$, so
$SU(3)_3$ is an open neighbourhood of the identity. The inverses of
the local trivializations are given by

$\psi_3\colon S^5_3\times SU(2)\to SU(3)_3$, $\left(\pmatrix{a\cr
b\cr c\cr},R_3\right)\mapsto CR^\prime_3$, 

$\psi_2\colon S^5_2\times SU(2)\to SU(3)_2$, $\left(\pmatrix{a\cr
b\cr c\cr},R_2\right)\mapsto BR^\prime_2$, 

and

$\psi_1\colon S^5_1\times SU(2)\to SU(3)_1$, $\left(\pmatrix{a\cr
b\cr c\cr},R_1\right)\mapsto AR^\prime_1$

\noindent 
with $\psi_i=\varphi^{-1}_i$ after identifying $SU(2)\cong
SU(2)^\prime$, and $R^\prime_k=\pmatrix{
  &  &0 \cr
  &  &0 \cr
0 &0 &1 \cr}$,\hglue-47.5truept \raise6pt\hbox{\bigggfntit
R$_k$}\hglue47.5truept 
$k=1,2,3$. These formulae give {\it 
all\/} elements of $SU(3)$ in terms of points of the $3$- and
$5$-spheres. 

\

With the help of the above formulae the set of matrices of $SU(3)$ can 
be divided into seven disjoint subsets: $SU(3)_{123}$, $SU(3)_{i,jk}$ 
and $SU(3)_{ij,k}$, respectively the pieces of $SU(3)$ lying over $S^5_
{123}=S^5_1 \cap S^5_2 \cap S^5_3$, $S^5_{i,jk}=S^5\setminus 
(S^5_j \cup S^5_k)$ 
and $S^5_{ij,k}=S^5_i \cap S^5_j \setminus 
S^5_{123}$, with $i,j,k \in \{1,2,3 \}$ 
in cyclic order:
$$\pmatrix{{\vert c \vert ^2 z/c +a \bar b
\bar w \over \sqrt{1-\vert b \vert ^2}} & {\vert c \vert^2 w/c-a \bar b \bar z 
\over \sqrt {1-\vert b \vert ^2}} & a \cr
-\bar w \sqrt {1-\vert b \vert ^2} & \bar z \sqrt {1-\vert b \vert ^2} & b \cr 
{-\bar a z +\bar b c \bar w \over \sqrt {1-\vert b \vert ^2}} & 
-{\bar a w+\bar b c \bar z \over \sqrt {1-\vert b \vert ^2}} 
& c \cr}\in SU(3)_{123}, $$
$\vert a \vert ^2+\vert b \vert ^2+\vert c \vert ^2=1$ ,$0<\vert a \vert , 
\vert b \vert , \vert c \vert <1$; 
$$\pmatrix {0 & 0 & e^{i\varphi } \cr
ze^{-i\varphi} & we^{-i\varphi} & 0 \cr
-\bar w  & \bar z & 0 \cr}\in SU(3)_{1,23};$$
$$\pmatrix {-ze^{-i\varphi} & 
-we^{-i\varphi} & 0 \cr 0 & 0 & e^{i\varphi} \cr 
-\bar w & \bar z & 0 \cr}\in SU(3)_{2,31}; $$
$$\pmatrix {ze^{-i\varphi} & we^{-i\varphi} & 0 \cr 
-\bar w & \bar z & 0 \cr 0 & 0 & 
e^{i\varphi} \cr}\in SU(3)_{3,12}; $$
$$\pmatrix{-z\bar b & 
-w\bar b & a \cr z\vert a \vert^2 \over a & w\vert a \vert^2 \over a & b \cr
-\bar w & \bar z & 0 \cr}\in SU(3)_{12,3},$$
$\vert a \vert ^2+ \vert b \vert ^2=1$, $a, b \neq 0$;
$$\pmatrix {-z\vert b \vert \over b & -w\vert b \vert \over b 
& 0 \cr \bar w b\bar c \over \vert b \vert & -\bar z b\bar c \over \vert b \vert 
& b \cr -\vert b \vert \bar w
& \vert b \vert \bar z & 
c \cr}\in SU(3)_{23,1}, $$
$\vert b \vert ^2+\vert c \vert ^2=1$, $b,c \neq 0$;
$$\pmatrix {\vert c \vert ^2 z \over c & \vert c \vert ^2 w \over c & a \cr 
-\bar w & \bar z & 0 \cr -\bar az & -\bar aw & c \cr}\in SU(3)_{31,2},$$
$\vert a \vert ^2+\vert c \vert ^2=1$, $a,c \neq 0$; with $\vert z \vert ^2
+\vert w \vert ^2=1$ and $\varphi \in [0,2\pi )$.

\

{\it Remark.} The above results can be extended to the bundles 
$$U(n-1) \to U(n) \aderecha{p_n}\ S^{2n-1}$$
{\it i.e.} to the {\it unitary groups}. In particular for $n=3$ we have 
the pair of maps
$$U(2) \aderecha{\iota}\ U(3) \aderecha{p_3}\ S^5$$
with $U(2)=\bigg \{ \pmatrix {z & w \cr \bar we^{i\lambda} & -\bar ze^{i
\lambda} \cr} \bigg \vert \vert z \vert^2+ \vert w \vert^2=1,\lambda \in 
[0,2\pi) \bigg \}$. The local trivializations of $U(3)$, $\phi_k: U(3)_k 
\to S^5_k \times U(2)'$, $k=1,2,3$ and where $U(2)'=\iota (U(2))$ are 
given by the same formulae as those for $SU(3)$, with the matrices $A$, 
$B$ and $C$ respectively replaced by the matrices $\hat{A}$, $\hat{B}$ 
and $\hat{C}$.

\section{3. $SU(3)$ from the $n=2$ Hopf bundle} 

\subsection{3.1. Suspension}

\

The {\it suspension} of a topological space $X$ is the quotient space given  
by 
$$SX={X \times I \over {X\times \{0\},X\times \{1\}}}=\{[x,t]\}_{(x,t)\in X 
\times I}$$
with 
$$[x,t]=\bigg\{\matrix{\{(x,t)\},t \in (0,1) \cr
X\times \{0\}, t=0 \cr X\times \{1\}, t=1. \cr}$$
This means that in the product $X\times I$, $X\times 
\{0\}$ has been identified to one point, and $X\times \{1\}$ has been 
identified to another point. Intuitively, it is clear that $SS^0\cong S^1$, 
$SS^1\cong S^2$, ..., $SS^{n-1}\cong S^n$. The suspension of a continuous  
function is defined by $Sf([x,t])=[f(x),t]$, which satisfies the {\it 
functorial} properties $Sid_X=id_{SX}$ and $S$ $g\circ f=Sg\circ Sf$ if 
$f:X\to Y$ and $g:Y \to Z$. If $p_Z:Z \times I \to SZ$ is the projection 
$p(z,t)=[z,t]$ then the following diagram commutes:
$$\matrix{X & \aderecha{f}\ & Y \cr \iota_0 \downarrow & & \downarrow 
\iota_0 \cr X \times I & \aderecha{f \times id}\ & Y \times I \cr 
p_X \downarrow & & \downarrow p_Y \cr SX & \aderecha{Sf}\ & SY \cr}.$$
If $H:X \times I \to Y$ is 
a homotopy between $h_0$ and $h_1$ then $SH:SX \times I 
\to  SY$ given by $SH([x,t],t^\prime)=[H(x,t^\prime),t]$ {\it i.e.} 
$(SH)_{t^\prime}=SH_{t^\prime}$ is a homotopy between $Sh_0$ and $Sh_1$.  
$SH$ is called the {\it suspension of the homotopy}. Then there is a well 
defined function between homotopy classes of maps $S:[X,Y] \to [SX,SY]$, 
$[f] \to S([f]):=[Sf]$.

\

If $X$ is a pointed space with base point $x_0$, then the 
{\it reduced suspension} of $X$, $S_rX$ is defined by
$$S_rX={X \times I \over X \times \{0\} \cup X \times \{1\} \cup \{x_0\}
\times I}$$ {\it i.e.} all the points in $X \times \{0\}$, $X \times 
\{1\}$ and $\{x_0\} \times I$ are identified to one point. In this case its 
elements are given by 
$$[x,t]= \bigg\{\matrix{X \times \{0\} \cup X \times \{1\} \cup \{x_0\} 
\times I, \matrix{x=x_0,\ all \ t \in I \cr t=0 \ or \ 1,\ all \ x \in X 
\cr} \cr \{(x,t)\}, t \in (0,1) \ and \ x \neq x_0. \cr}$$
$\tilde{x}_0=[x_0, t]$ is the base point of $S_rX$. If $f:X \to Y$  
preserves the base points {\it i.e.} if $f(x_0)=y_0$, then $S_rf(\tilde{x}
_0)=y_0$, and if $h_0 \matrix{H \cr \sim \cr}h_1 \ (rel \ x_0)$ then 
$Sh_0 \matrix{SH \cr \sim \cr}Sh_1 \ (rel \ \tilde{x}_0)$. ($rel \ x_0$ 
means that the homotopy $H$ preserves the base point.) Also, there 
is a homeomorphism $\varphi^{-1}_{n+1}:S_rS^n \to S^{n+1}$ given by
$$\varphi ^{-1}_{n+1}([\vec {x},t])=\bigg\{\matrix{p^{-1}_-(2t\vec {x}+(1-2t)
\vec{x}_0), \ t \in [0,1/2] \cr p^{-1}_+((2-2t)\vec{x}+(2t-1)\vec{x}_0), 
\ t \in [1/2,1] \cr}$$ 
where : $\vec {x}_0=(1,0,...,0) \in S^{n+1}=\{(x_1,...,x_{n+2}) \vert 
\sum\limits^{n+2}_{i=1} x^2_i=1 \} \subset \R^{n+2}$ is the base point, 
$S^n=\{\vec {x}\in S^{n+1} \vert x_{n+2}=0\}$, and if $H_+=\{\vec{x} 
\in S^{n+1} \vert x_{n+2} \geq 0\}$, $H_-=\{\vec {x} \in S^{n+1} 
\vert x_{n+2} \leq 0\}$ and $D^{n+1}=\{(x_1,...,x_{n+1},0) \vert 
\sum\limits ^{n+1}_{i=1} x_i^2 \leq 1 \}$ then $p_{+(-)}
:H_{+(-)} \to D^{n+1}$ are the homeomorphisms given by $(
x_1,...,x_{n+2})\mapsto (x_1,...,x_{n+1},0)$ with inverses $(x_1,..., 
x_{n+1},0) \mapsto (x_1,...,x_{n+1},+(-) \sqrt{1-\sum
\limits ^{n+1}_{i=1} x_i^2 })$ respectively (Spanier, 1966). In particular, 
$$\tilde{\vec{x}}_0=S^n \times \{0\}=S^n \times \{1\}=\{\vec{x}_0\} \times I
\aderecha{\varphi^{-1}_{n+1}}\ \vec{x}_0$$ and if $\vec{x}\neq \vec{x}_0$ 
then $[\vec{x},1/2]=\{(\vec{x},1/2)\}\aderecha{\varphi^{-1}_{n+1}}\ \vec{x}.$ 

\

The inverse homeomorphism is given by the following formulae: $\vec{x}_0 
\mapsto \tilde {\vec{x}}_0$, if $\vec {x} \in S^n$ and $\vec{x} \neq 
\vec{x}_0$ then $\vec{x} \mapsto [\vec{x},1/2]=\{(\vec{x},1/2)\}$, 
$(0,...,0,1)=N \ (north \ pole) \mapsto [-\vec{x}_0,3/4]=
\{(-\vec{x},3/4)\}$, 
$(0,...,-1)=S \ (south \ pole) \mapsto [-\vec{x}_0,1/4]=
\{(-\vec{x}_0,1/4)\}$, and if        
$\vec{x} \in H_{+(-)} \ \setminus \ S^n$, $\vec{x} \neq S,N$, then 
$$\varphi_{n+1}(\vec{x})=[\vec{z}(\vec{x}),t_{+(-)}(\vec{x})]=
\{(\vec{z}(\vec{x}),t_{+(-)}(\vec{x}))\}$$ with 
$$\vec{z}(\vec{x})={(2(1-x_1)x_1-x^2_{n+2},
2(1-x_1)x_2,...,2(1-x_1)x_{n+1}) \over {2(1-x_1)-x^2_{n+2}}}$$ 
and $t_{+(-)}(\vec{x})=1/2+(-){x^2_{n+2}\over {4(1-x_1)}}$.

\subsection{3.2. $SU(3)$ from the Hopf map $h$}

\

Let $G$ be a path connected topological group. Then the set of 
isomorphism classes of principal $G$-bundles over the $n$-sphere 
$k_G(S^n)$ is in one-to-one correspondence with $\Pi _{n-1}(G)$ 
(Steenrod, 1951). This can be understood from the fact that the 
$n$-sphere can be covered by two open sets $U_1, U_2$, which are 
homeomorphic to $n$-balls and contain $S^{n-1}$, and the fact that 
any bundle over an $n$-ball is trivial. Using these trivializations 
there is only one transition function $g_{12}:U_1 \cap U_2 \to G$, for a 
bundle $\xi$. Then we associate to $\xi$ the map $g_{12}\mid _
{S^{n-1}}:S^{n-1}\to G$, called the characteristic map of $\xi$. 
Therefore, if the characteristic maps of two bundles are in the same 
homotopy class, then the corresponding bundles are 
isomorphic, and a bundle is trivial if and only if its characteristic 
map is null-homotopic. Notice that in our construction of 
the local charts for $SU(3)$ we have used a different trivialization. In 
the following we shall consider the bundles $SU(n-1) \to SU(n) \aderecha
{\pi_n}\ S^{2n-1}$ and call $g_n:S^{2n-2}\to SU(n-1)$ 
the corresponding characteristic maps.

\

To study the case $n$=3 we need the following

\

{\it Proposition.} The successive suspensions of the Hopf map, $S_rh:
S^4 \to S^3$, $S^2_rh:S^5\to S^4$,... are essential (Steenrod and 
Epstein, 1962).

As a consequence, we have the 

\

{\it Proposition.} $SU(3)$ is determined by the suspension of the Hopf map. 

\

{\it Proof.} For $n=3$, $k_{SU(2)}(S^5) \cong [S^4,S^3] \cong \Pi_4(S^3) \cong 
\Z_2=\{0,1\}$ and $g_3:S^4 \to S^3$. 
By the proposition above $S_r h$ is essential. To see that 
$g_3$ is also essential we will show that the bundle $SU(3) \aderecha{
\pi _3} S^5$ is not trivial. By  
(EDM, 1993), $\Pi _4(SU(3)) \cong 0$, on the other hand 
$\Pi _4(S^5 \times SU(2)) \cong 
\Pi _4(S^5)\times \Pi _4(S^3) \cong 0 \times \Z_2 \cong \Z _2$. Hence $SU(3)$ is not 
isomorphic to the trivial bundle. Since $\Pi _4(S^3) \cong \Z _2$ we have 
that $[g_3]=[S_r h]$.   QED

\section{4. The case of $SU(n)$}

\subsection{4.1. $SU(4)$}

\

For $n=4$, $k_{SU(3)}(S^7)\cong [S^6,SU(3)]\cong \Pi_6(SU(3))\cong 
\Z_6$ which has two generators. 
This means that up to isomorphism there are five nontrivial $SU(3)$-
bundles over $S^7$, one of them being $SU(4)$ since $\Pi _6(SU(4)) \cong 
0$ and $\Pi _6(S^7 \times SU(3)) \cong \Pi _6(S^7) \times \Pi _6(SU(3)) 
\cong \Z_6$. Let $g_4:S^6 \to SU(3)$ be 
its characteristic map. If $g_4$ were a homotopy lifting of $S^3_rh$, 
then there should exist a lifting $g^\prime_4$ by 
$\pi_3:SU(3) \to S^5$ of $S_r^3h :S^6 \to S^5$ {\it i.e.} a 
commuting diagram 
$$\matrix{ & & SU(3) & & \cr & {}^{g^\prime_4} \nearrow & & \searrow 
{}^{\pi_3} & \cr S^6& & \aderecha{S^3_rh}\ & & S^5 \cr}$$ 
with $g_4^\prime \sim g_4$. We have the

\

{\it Proposition.} $\pi_3$ does {\it not} lift $S^3_rh$.

\

{\it Proof.} We will show that the homomorphism $\pi_{3*}:\Pi_6(SU(3))\to
\Pi_6(S^5)$ is zero. This implies that any map $S^6 
\aderecha{f}\ S^5$ which factorizes through $\pi_3$ {\it i.e.} a map   
for which there exists a map $S^6 \aderecha{g}\ SU(3)$ such that $\pi_3 
\circ g\sim f$, is null-homotopic. The result now follows from this since 
$S^3_rh$ is essential. 

\

Consider the long exact homotopy sequence (Steenrod, 1951) of the principal 
bundle $SU(2)\to SU(3) \aderecha{\pi_3}\ S^5$:
$$...\to \Pi_6(SU(3))\aderecha{\pi_{3*}}\ \Pi_6(S^5) \aderecha{\delta}\ 
\Pi_5(S^3)\aderecha{\iota_*}\ \Pi_5(SU(3)) \to ...$$
This gives an exact sequence:
$$...\to \Z_6 \aderecha{\beta}\ \Z_2 \aderecha{\gamma}\ \Z_2 
\aderecha{\alpha}\ \Z \to...$$ where we called $\beta$, $\gamma$ and 
$\alpha$ the homomorphisms corresponding to $\pi_{3*}$, $\delta$ and 
$\iota_*$, respectively. Since $\Z_2$ is a torsion group and $\Z$ is 
torsion free, then the homomorphism $\alpha $ is zero. Therefore 
$\gamma$  is an isomorphism 
{\it i.e.} $ker(\gamma )=ker(\delta )=\{0\}=Im(\pi_
{3*})$ {\it i.e.} $\pi_{3*}=0.$   QED

{\it Remark.} This result can also be obtained from the general theorem proved in Section 4.4.  However, the proof given above is simpler.

\subsection{4.2. $(H,f)$-structures}

\

Let $H$ and $G$ be topological groups (e.g. Lie groups), $\xi _H:H \to 
PH \aderecha{\pi _H} BH$ and $\xi _G:G \to PG \aderecha{\pi _G} BG$ their 
universal bundles, and $f:H \to G$ a topological group 
homomorphism. Then the action $\bar f:H\times G \to G$, $\bar f(h,g) =f(h)g$ 
induces the associated principal $G$-bundle $(\xi _H)_G: G \to PH \times _H 
G \to BH$ with total space $PH \times _H G= \{[a,g]\}_
{(p,g)\in PH \times G}$, $[a,g]=\{(ah,f(h^{-1})g)\}_{h\in H}$, 
action $(PH \times _H G)\times G \to PH\times _H G$ given by $[a,g]\cdot 
g^\prime =[a,gg^\prime ]$, and projection $(\pi_H)_G([a,g])=\pi_H(a)$.  
$PH\times _H G$ is isomorphic to the pull-back 
bundle $(Bf)^*(PG)$, where the induced function $Bf:BH \to BG$ is uniquely 
defined up to homotopy. 

\

If $HTop$ is the category of paracompact topological spaces and homotopy 
classes of maps, and $Set$ is the category of sets and functions, then  
for each topological group $K$ there are two cofunctors $k_K$ and $[$  $,BK]$ 
from $HTop$ to $Set$ such that, for each topological group 
homomorphism $f:H \to G$ there are natural transformations $f_*:
k_H \to k_G$ and $Bf_*:[$  $,BH] \to [$  $,BG]$, and natural equivalences 
$\psi _H$ and $\psi _G$ which make the following functorial diagram 
commutative: $$\matrix{k_H & \aderecha{f_*} & k_G\cr \psi _H \uparrow & & 
\uparrow \psi _G \cr [$  $,BH] & \aderecha{Bf_*} & [$  $,BG] \cr }$$
So, for each paracompact topological space $X$ the following set theoretic diagram 
commutes: $$\matrix{k_H(X) & \aderecha{f_*} & k_G(X) \cr \psi_H \uparrow \cong
& & \cong \uparrow \psi_G \cr [X,BH] & \aderecha{Bf_*} & [X,BG] \cr }$$
where: $k_K(X)$= $\{$isomorphism classes of principal $K$-bundles over $X \}$, $
[X,BK]$= $\{$ unbased homotopy classes of maps from $X$ to $BK \}$, $\psi _K (
[\alpha ])=[\alpha ^*(PK)]$, $f_*([\eta ])=[\xi]$ with $\eta :H \to E 
\aderecha{q} X$ and $\xi :G \to E \times _H G \aderecha{\bar q} X$, and
$Bf_*([\alpha ])=[Bf \circ \alpha]$. If $[\xi ] \in k_G(X)$, then 
$f_*^{-1}(\{[\xi ] \})$ is the set of $(H,f)$-{\it structures} on $\xi $; 
this set can be empty. So, $\xi $ has a $(H,f)$-structure if and only if 
there exists a map $\alpha :X \to BH$ such that $Bf \circ \alpha \sim F$, 
where $F$ is the classifying map of $\xi $. One can show that this 
definition is equivalent to the existence of a $G$-bundle isomorphism 
$$\matrix{
E \times _H G   & \aderecha{\bar {\varphi }} & P \cr
\bar q \searrow &                            & \swarrow \pi \cr
                & X                          & \cr }$$
where $H \to E \aderecha{q} X$ is a
principal $H$-bundle or, equivalently,  
to the bundle map
$$\matrix{
E \times H        & \aderecha{\varphi \times f} & P \times G \cr
\kappa \downarrow &                             & \downarrow \psi \cr
E                 & \aderecha{\varphi}          & P \cr
q \searrow        &                             & \swarrow \pi \cr
                  & X                           &  \cr}$$
where $\varphi
=\bar {\varphi }\circ \varphi _f$ with $\varphi _f:E \to E\times _H G$ given by $
\varphi _f(a)=[a,e]$ ($e$ is the unit of $G$) (Aguilar and Socolovsky, 1997a). One  
says that $(E,\varphi)$ {\it is an} $(H,f)$-{\it structure on} $G \to P 
\aderecha{\pi } X$.

\

In the case of smooth bundles, if the Lie group homomorphism $H \aderecha{f} G$ 
is an {\it embedding} {\it i.e.} an 
injective immersion, then $E$ is called a {\it reduction} of $P$ to $H$. 
In this setting, one has the following

\

{\it Proposition.} If $f$ is an embedding, then $\varphi $ is also an 
embedding. 

\

{\it Proof.} Since $\varphi =\bar \varphi \circ \varphi _f$ and $ \bar 
\varphi $ is a diffeomorphism, then $\varphi $ is an embedding if and only 
if $\varphi _f$ is an embedding; we shall show that $\varphi_f$ is an  
embedding. i) $\varphi _f$ 
is injective: Let $\varphi _f (a_1)=\varphi _f (a_2)$ {\it i.e.} $[a_1,e]=
[a_2,e]$, since $[a,e]=\{(ah,f(h^{-1}))\}_{h \in H}$ then there must 
exist $h \in H$ such that $(a_1,e)=(a_2 h,f(h^{-1})$ {\it i.e.} 
$a_1=a_2 h$ and $f(h^{-1})=e$, but $f$ is injective, so $h^{-1}=h=e^\prime$,  
the identity in $H$, and then $a_1=a_2$. ii) 
$d\varphi _f$ is injective at each $a_0 \in E$:  
Consider the commutative diagram
$$\matrix{
  &        & E\times G             &             \cr
  &i\nearrow &                       & \searrow p   \cr
E &        & \aderecha{\varphi _f} & E\times _H G \cr}$$
where $i(a)=(a,e)$ and $p(a,e)=[a,e]$. By (Greub et al, 1973) $H \to$
$E\times G \aderecha{p} E\times _H G$ is a principal $H$-bundle, so fixing $(a_0,e) 
\in E\times G$ there is a map $\alpha _{(a_0,e)}:H \to$ 
$E \times G$, given by $\alpha _{(a_0,e)}(h)=(a_0,e)\cdot h=(a_0h,f(h^{-1}))$, in 
particular 
$\alpha _{(a_0,e)}(e^\prime)=(a_0,e)$. One then has the following diagram 
of vector spaces:
$$\matrix{0 \to 
T_{e^\prime }H \aderecha{(d\alpha _{(a_0,e)})_{e^\prime}}
 & T_{(a_0,e)}(E\times G) \aderecha{(dp)_{(a_0,e)}}
 & T_{[a_0,e]} (E\times _H G) \to 0 \cr\noalign{\vskip3pt}
& (di)_{a_0} \uparrow
 & \nearrow (d\varphi _f )_{a_0} &  \cr\noalign{\vskip3pt}
& T_{a_0}E   & &\cr}$$
where the horizontal sequence is exact and the triangle commutes. 
If $p_1$ and $p_2$ are respectively the 
projections of $E\times G$ onto $E$ and $G$, then  
$p_1 \circ i (a)=a$ {\it i.e.} $p_1 \circ i=id_E$ and $p_2 \circ i(a)=p_2
(a,e)=e$ {\it i.e.} $p_2 \circ i=const.$, hence $(d(p_1 \circ i))
_{a0}=(d(id_E ))_{a_0}=id_{T_{a_0}E}$ and $(d(p_2 \circ i))_{a_0}=
(d(const.))_{a_0}=0$.  Therefore $(di)_{a_0}(v)=
((d(p_1 \circ i))_{a_0} (v),(d(p_2 \circ i))_{a_0} (v))=(v,0)$.  
On the other hand, $p_1 \circ \alpha_{(a_0,e)}(h)=p_1(a_0 h,f(h^{-1}))=a_0 
h:=\alpha _{a_0}(h)$, $p_2 \circ 
\alpha_{(a_0,e)}(h)=p_2(a_0 h,f(h^{-1}))=f\circ \gamma (h)$, where $\gamma 
:H \to H$ is given by $\gamma (h)=h^{-1}$. Therefore 
$(d\alpha _{(a_0,e)})_{e^\prime}
(w)=((d(p_1 \circ \alpha _{(a_0,e)}))_{e^\prime}(w),(d(p_2 \circ \alpha _
{(a_0,e)}))_{e^\prime}(w))=((d\alpha_{a_0})_{e^\prime}(w),(df)_{e^\prime}
\circ (d\gamma)_{e^\prime}(w))$. Let $(r,s) \in Im((di)_{a_0})
\cap Im((d\alpha _{(a_0,e)})_{e^\prime}))$, then $s=0$ and hence $0=(df)_{
e^\prime}((d\gamma)_{e^\prime}(w))$, therefore $(d\gamma)_{e^\prime}(w)=0$ 
because $f$ is an immersion and, since $\gamma $ is a diffeomorphism, 
$w=0$, so $r=(d\alpha_{a_0})_{e^\prime}(0)=0$ {\it i.e.} $Im((di)_{a_0})
\cap Im((d\alpha_{(a_0,e)})_{e^\prime})=\{0\}$. Finally, let $v \in  
ker ((d\varphi _f)_{a_0})$, then $0=(dp)_{(a_0,e)}((di)_{a_0})(v))$ {\it 
i.e.} $(di)_{a_0}(v)\in ker((dp)_{(a_0,e)})=Im((d\alpha_{(a_0,e)})_
{e^\prime})$ {\it i.e.} $(di)_{a_0}(v)=0$. Since $i$ is an embedding, 
then $v=0$ {\it i.e.} $(d\varphi _f)_{a_0}$ is one-to-one.   QED

\

{\it Remark.} One often finds in the literature (Kobayashi and Nomizu, 1963; 
Trautman, 1984) that to define a reduction to a Lie subgroup $H\subset G$, 
$\varphi$ is required to be an embedding. The proposition above shows 
that this is a consequence of the fact that $H\to G$ is an embedding.

\subsection{4.3. $SU(n)\to SU(n+1)\aderecha{\pi_{n+1}} S^{2n+1}$ as an 
$(SU(n),\iota)$-structure on $U(n) \to U(n+1) \aderecha{p_{n+1}} S^{2n+1}$}

\

{\it Proposition.} For $n=1,2,3,...$, the bundle $\pi_{n+1}$  is a reduction of 
the bundle $p_{n+1}$ {\it i.e.} one has the $U(n)$-bundle isomorphism given by the commutative diagram 
$$\matrix{
(SU(n+1)\times _{SU(n)} U(n))\times U(n) & \aderecha{\bar {\varphi }\times id} & U(n+1)\times U(n) \cr
\lambda \downarrow                       &                                   & \downarrow \psi \cr
SU(n+1)\times _{SU(n)} U(n)              & \aderecha{\bar {\varphi}}           & U(n+1)  \cr
q_{n+1} \searrow                         &                                   & \swarrow p_{n+1} \cr
                                         & S^{2n+1}                          & \cr}$$

\noindent
where $\lambda ([D,A],B)=[D,AB], q_{n+1} [D,A]=\pi_{n+1} (D)=De_0, \psi 
(C,B)=Cj(B), \break p_{n+1}(C)=Ce_0$, and $\bar {\varphi}$, and $j$ 
are given below.

\

{\it Proof.} Consider the inclusion $SU(n+1) \aderecha{\varphi} U(n+1)$; one 
can easily show that this is a smooth bundle map between the principal 
$SU(n)$-bundle $\pi_{n+1}$ and the principal $U(n)$-bundle $p_{n+1}$. 
Therefore, by (Greub et al, 1973) the map $\bar \varphi$ given by  
$\bar \varphi ([D,A])=Dj(A)$ where $j$ is the inclusion $U(n)\to U(n+1)$ 
with $j(A)=\pmatrix{A & 0 \cr 0 & 1 \cr}$, is a smooth bundle isomorphism. 
The inverse of $\bar \varphi$ is given as follows: if $C \in 
U(n+1)$ then $C=Dl(det C)$ where $D=C(l(det C))^{-1} \in SU(n+1)$ and $
l:U(1) \to U(n+1)$ is the inclusion $l(z)=\pmatrix{z & 0 \cr 0 & I \cr}$, 
then $[D,l(det C)]= \bar \varphi ^{-1}(C)$. QED

{\it Remark}. Notice that if $\iota:SU(n) \to U(n)$ is the inclusion, then 
$p_n \circ \iota=\pi_n$. 

\subsection{4.4. Proof of the main result}

\

{\it Proposition.} Let $H$ and $G$ be path connected topological groups such that each one has the homotopy type of a CW-complex and let $f:H \to G$ be a continuous homomorphism. Then, in the following diagram each horizontal function is a bijection and each square commutes, for $n=1,2,3,...$
$$\matrix{
[S^{2n},G] & \aderecha{\mu _{G \#}} & [S^{2n},\Omega BG] & \aderecha{adj_{G*}} & 
[S_rS^{2n},BG] & \aderecha{\psi _G} & k_G(S^{2n+1}) \cr 
\uparrow f_{\#} & & \uparrow \Omega Bf_{\#} & & \uparrow Bf_* & & \uparrow f_* \cr 
[S^{2n},H] & \aderecha{\mu _{H \#}} & [S^{2n},\Omega BH] & \aderecha{adj_{H*}} &
[S_rS^{2n},BH] & \aderecha{\psi _H} & k_H(S^{2n+1}) \cr}$$

\noindent
where $k_K$, $f_*$, $Bf_*$ and $\psi _K$ have been defined before, $\Omega BK$ is the loop space of $BK$, and $f_{\#}$, $\mu _{K \#}$, $\Omega Bf_{\#}$ and $adj_{K*}$ are given by $f_{\#}([\delta ])=[f \circ \delta ]$, $\mu _{K \#}([\sigma ])=[\mu _K \circ \sigma ]$ ($\mu _K$ is defined below), 
$\Omega Bf_{\#}([\kappa ])=[\Omega Bf \circ \kappa]$ with $\Omega Bf:\Omega BH \to \Omega BG$ given by $\Omega Bf(\gamma )=Bf \circ \gamma$, and $adj_{K*}([\alpha ])=[adj_K(\alpha )]$ with $adj_K(\alpha )([z,t])=\alpha (z)(t)$, $t \in [0,1]$. The set $[S^{2n},K]=\Pi_{2n}(K)$ corresponds to the characteristic maps for the $K$-principal bundles over $S^{2n+1}$. 

\

{\it Proof.} The commutativity of the third square has been proved in 
Section 4.2, with $S^{2n+1}=X$. The natural equivalence $adj_K$ is given 
by the {\it exponential law} in function spaces (Spanier, 1966). By (Switzer, 1975) there exist {\it homotopy equivalences} $\mu _K$ 
such that the diagram 
$$\matrix{
H               & \aderecha{f}         & G \cr
\mu _H \uparrow &                      & \uparrow \mu _G \cr
\Omega BH       & \aderecha{\Omega Bf} & \Omega BG \cr}$$
commutes up to homotopy, therefore $\mu _{K \#}$ is a bijection and the first square commutes. Finally, notice that in the diagram of the proposition we are dealing with based homotopy classes of maps. This corresponds to based principal bundles. However, since we are taking path connected topological groups, the function that forgets the basepoints is a bijection between based bundles and the usual unbased bundles of section 4.2. \hskip 1cm   QED

\

{\it Proposition.} For even $n$, $n \geq 2$, the clutching map $g_{n+1}$ of 
the principal bundle $SU(n) \to SU(n+1) \aderecha{\pi _{n+1}} S^{2n+1}$ is 
a homotopy lifting of the $(2n-3)-th$ reduced suspension of the Hopf map $h$. 
For odd $n$, $n \geq 3$, $\pi_n \circ g_{n+1}$ is inessential.

\

{\it Proof.} We apply the previous proposition to the case $H=SU(n)$, 
$G=U(n)$, and $f=\iota$ (the inclusion), for $n=2,3,...$. By the proposition  
in Section 4.3, $[\xi ]=\iota _*([\eta ])$ with $\xi :U(n) \to U(n+1)  
\aderecha{p_{n+1}} S^{2n+1}$ and $\eta :SU(n) \to SU(n+1) \aderecha{\pi 
_{n+1}} S^{2n+1}$. Then one has the commutative diagram 
$$\matrix{
[T^\prime _{n+1}] & \in & [S^{2n},U(n)]        & \aderecha{\mu } &k_{U(n)}(S^{2n+1})   & \ni & [\xi ] \cr
                  &     & \iota _{\#} \uparrow &                 &\iota _* \uparrow    &     &        \cr
[g_{n+1}]         & \in & [S^{2n},SU(n)]       & \aderecha{\nu } & k_{SU(n)}(S^{2n+1}) & \ni & [\eta ] \cr}$$
where $\mu =\psi _{U(n)} \circ adj_{U(n)*} \circ \mu _{U(n) \#}$ and 
$\nu =\psi _{SU(n)} \circ adj_{SU(n)*} \circ \mu _{SU(n) \#}$ are bijections, 
and $T^\prime _{n+1}$ is the clutching map for the principal bundle $\xi $ 
(Steenrod, 1951). Then $[T^\prime _{n+1}]=\mu ^{-1}([\xi ])=\mu ^{-1}(
\iota _*([\eta ]))=\mu ^{-1} \circ \iota _* (\nu ([g_{n+1}]))=
\mu ^{-1} \circ \iota _* \circ \nu ([g_{n+1}])=\iota _{ \#}([g_{n+1}])=
[\iota \circ g_{n+1}]$ and therefore $T^\prime _{n+1} \sim \iota \circ 
g_{n+1}$. 

\

Consider the following diagram: 
$$\matrix{
       &                          & U(n-1)                &              & \cr
       &                          & \downarrow            &              & \cr
       &                          & U(n)                  &              & \cr
       & T^\prime _{n+1} \nearrow &                       & \searrow p_n & \cr
S^{2n} &                          &\aderecha{S_r^{2n-3}h} &              &S^{2n-1} \cr}$$
Steenrod (Steenrod, 1951) proved that for $n$ even, $n \geq 2$, $p_n 
\circ T^\prime _{n+1} \sim S_r^{2n-3}h$ {\it i.e.} the diagram commutes 
up to homotopy, while for $n$ odd, $n \geq 3$, $p_n \circ T^\prime _
{n+1} \sim const.$ Then, $p_n \circ T^\prime _{n+1} \sim p_n \circ 
(\iota \circ g_{n+1})=(p_n \circ \iota )\circ g_{n+1}=\pi _n \circ 
g_{n+1}$
$$
\sim
\bigg\{ \matrix{S_r^{2n-3}h, \ n \ even \cr const., \ n \ odd \cr}
$$
QED

\section{5. $S^2$ and relativity}

\

As is well known, the Lorentz group, the group of linear 
transformations of Minkowski space-time which preserves the scalar 
product $<x,y>=x^T \eta y$ where $\pmatrix{1 & 0 & 0 & 0 \cr 
0 & -1 & 0 & 0 \cr 
0 & 0 & -1 & 0 \cr 0 & 0 & 0 & -1 \cr}$ is the Minkowskian metric, 
is a subgroup of 
the symmetry group of several gauge theories of gravity (Hehl {\it et 
al}, 1976; Basombr\'\i o, 1980). This means that $O(3,1)$ is a 
subgroup of the structure group of the corresponding principal bundles. 
The relationship between these theories and the 2-sphere (the Riemann 
sphere $\C \cup \{ \infty \}$) comes from the fact that there is a 
canonical isomorphism between the connected component of $O(3,1)$, the 
proper orthochronous Lorentz group $SO^0(3,1)$ and the group of 
conformal (Moebius) transformations of $S^2$, $Conf(S^2)$. We recall 
that $Conf(S^2)$ is the set of all invertible transformations of the 
Riemann sphere which preserves the angles between curves and that at 
each point multiply all the tangent vectors by a fixed positive number. 

Let $g=\pmatrix{a & b \cr c & d \cr}$ be an element of $GL_2(\C)$, we 
define a Moebius transformation $m:S^2 \to S^2$ as follows: if $c \neq 
0$, then 
$$
z \mapsto
\cases{{az+b\over cz+d} & if $z \neq -d/c$\cr
\infty & if $z= -d/c$\cr}
$$
and  
$$\infty \mapsto a/c;$$

and, if $c=0$, then 
$$
\left\{\matrix{z\mapsto {a\over d}z + {b\over d}\cr
\infty \mapsto \infty\cr}\right.
$$

It is then easy to verify that the following diagram commutes: 
$$\matrix{
          &               & \Z _2           &                  & \cr
          &               & \downarrow      &                  & \cr
          &               & SL_2(\C)        &                  & \cr
          & \psi \swarrow &                 & \searrow \lambda & \cr
SO^0(3,1) &               &\aderecha {\tau} &                  & Conf(S^2) \cr}$$
where: i) the projections $\psi $ and $\lambda $ are two-to-one 
group homomorphisms, respectively given by $\psi \pmatrix{a & b \cr 
c & d \cr}$ $$=\pmatrix{{\vert a \vert ^2 + \vert b \vert  
^2 + \vert c \vert ^2 + \vert d \vert ^2 \over 2} & Re(a \bar b+ c \bar d) & 
Im(a \bar b + c \bar d) & {\vert a \vert ^2 - \vert b \vert 
^2 + \vert c \vert ^2 - \vert d \vert ^2 \over 2} \cr Re(a \bar c + b \bar d) & 
Re(a \bar d + b \bar c) & Im(a \bar d - b \bar c) & Re(a \bar c- b 
\bar d) \cr -Im(a \bar c +b \bar d) & Im(a \bar d - b \bar c) & 
Re(a \bar d - b \bar c) & 
-Im(a \bar c - b \bar d) \cr {\vert a \vert ^2 + \vert b 
\vert ^2 - \vert c \vert ^2 - \vert d \vert ^2 \over 2} & Re(a \bar b -c \bar d) 
& Im(a \bar b - c \bar d) & {\vert a \vert ^2 - \vert b \vert  
^2 - \vert c \vert ^2 + \vert d \vert ^2 \over 2} \cr}$$
with $\psi \pmatrix{a & b \cr c & d \cr}\!=\psi \pmatrix{-a & -b \cr 
-c & -d \cr}\!=l$ (Penrose and Rindler, 1984), 
and $\lambda (g/ \sqrt {det g})
\!=m$ with $\lambda (g/ \sqrt {det g})= \lambda (-g/ \sqrt {det g})$; and 
ii) $\tau (l) = m$ is the desired isomorphism. $SL_2(\C) \aderecha{\psi } 
SO^0(3,1)$ and $SL_2(\C) \aderecha{\lambda } Conf(S^2)$ are $\Z_2$- principal 
bundles. 

\

Thus we conclude that the symmetry group of the standard model $G^\prime _
{SM}$, when gravitation is included, locally contains, as a space, $S^1 \times 
(S^3)^2 \times S^5 \times Conf(S^2)$. 

\

{\it Remark.} In the framework of the theory of categories, functors, and 
natural 
transformations, some of the geometrical objects of the previous 
sections, e.g. spheres and the Hopf map, have a natural origin. 
This suggests a possible relation between symmetries in nature, and 
therefore conservation laws, and some of the most general mathematical 
concepts. The basic idea is that of a representable functor (Aguilar 
and Socolovsky, 1997b).

\

\section{REFERENCES}

\

\noindent Aguilar, M. A., and Socolovsky, M. (1997a). Reductions and
extensions in bundles and homotopy, {\it Advances in Applied Clifford
Algebras}, {\bf 7 (S)}, 487-494.

\

\noindent Aguilar, M. A., and Socolovsky, M. (1997b). Naturalness of
the space of States in Quantum Mechanics, {\it International Journal of
Theoretical Physics}, {\bf 36}, 883-921.

\

\noindent Ashtekar, A., and Schilling, T.A. (1995). Geometry of Quantum 
Mechanics, {\it AIP Conference Proceedings}, {\bf 342}, 471-478.

\

\noindent Basombr\'\i o, F. G. (1980). A Comparative Review of Certain
Gauge Theories of the Gravitational Field, {\it General Relativity
and Gravitation}, {\bf 12}, 109-136.

\

\noindent Corichi, A., and Ryan, Jr., M. P. (1997). Quantization of 
nonstandard Hamiltonian systems, {\it Journal of Physics A: Mathematical 
and General}, {\bf 30}, 3553-3572.

\

\noindent Greub, W., Halperin, S. and Vanstone, R. (1973). {\it Connections, Curvature and Cohomology}, Academic Press, New York.

\  

\noindent Hehl, F. W., von der Heyde, P., and Kerlick, G. D. (1976).
General relativity with spin and torsion: Foundations and prospects,
{\it Reviews of Modern Physics}, {\bf 48}, 393-416.

\

\noindent It\^o, K. ed. (1993). {\it Encyclopedic Dictionary of
Mathematics}, The Mit Press, Cambridge, Mass. 

\

\noindent Kobayashi, S., and Nomizu, K. (1963). {\it Foundations of
Differential Geometry}, Vol. I, Wiley, New York.

\

\noindent Mohapatra, R. N. (1986). {\it Unification and
Supersymmetry}, Springer- Verlag, New York.

\

\noindent Penrose, R., and Rindler, W. (1984). {\it Spinors and
Space-time}, Vol. 1, Cambridge University Press, Cambridge.

\

\noindent Spanier, E. H. (1966). {\it Algebraic Topology},
Springer-Verlag, New York.

\

\noindent Steenrod, N. (1951). {\it The Topology of Fibre Bundles},
Princeton University Press, Princeton, New Jersey. 

\

\noindent Steenrod, N., and Epstein, D. B. A. (1962). {\it Cohomology
operations}, Annals of Mathematical Studies, {\bf 50}, Princeton
University Press, Princeton, New Jersey.

\

\noindent Switzer, R. (1975). {\it Algebraic Topology- Homotopy and
Homology}, Springer-Verlag, New York.

\

\noindent Taylor, J. C. (1976). {\it Gauge Theories of Weak
Interactions}, Cambridge University Press, Cambridge.

\

\noindent Trautman, A. (1984). {\it Differential Geometry for
Physicists}, Bibliopolis, Napoli. 

\

\noindent Wu, T. T., and Yang, C. N. (1975). Concept of nonintegrable 
phase factors and global formulation of gauge fields, {\it Physical 
Review D}, {\bf 12}, 3845-3857.

\end